# Instantaneous Three-dimensional Thermal Structure of the South Polar Vortex of Venus

I. Garate-Lopez (1), A. García Muñoz (2), R. Hueso (1, 3) and
A. Sánchez-Lavega (1, 3)

(1) Departamento de Física Aplicada I, E.T.S. Ingeniería, Universidad del País Vasco, Alameda Urquijo s/n, 48013 Bilbao, Spain.
(2) ESA Fellow, ESA/RSSD, ESTEC, 2200 AG Noordwijk, The Netherlands.
(3) Unidad Asociada Grupo Ciencias Planetarias UPV/EHU-IAA (CSIC) Bilbao, Spain.

Manuscript pages: 53
Figures: 12
Tables: 0

*Corresponding author address:*
Itziar Garate Lopez
Dpto. Física Aplicada I,
E.T.S. Ingeniería, Universidad del País Vasco,
Alda. Urquijo s/n, 48013 Bilbao, Spain
E-mail: itziar.garate@ehu.es
Telephone: (34) 94 601 7389
Fax: (34) 94 601 4178

**Abstract**

The Venus thermal radiation spectrum exhibits the signature of $CO_2$ absorption bands. By means of inversion techniques, those bands enable the retrieval of atmospheric temperature profiles. We have analyzed VIRTIS-M-IR night-side data obtaining high-resolution thermal maps of the Venus south polar region between 55 and 85 km altitudes. This analysis is specific to three Venus Express orbits where the vortex presents different dynamical configurations. The cold collar is clearly distinguishable centered at ~62 km (~100 mbar) altitude level. On average, the cold collar is more than 15 K colder than the pole, but its specific temperature varies with time. In the three orbits under investigation the South Polar Vortex appears as a vertically extended hot region close to the pole and squeezed by the cold collar between altitudes 55 and 67 km but spreading equatorwards at about 74 km. Both the instantaneous temperature maps and their zonal averages show that the top altitude limit of the thermal signature from the vortex is at ~80 km altitude, at least on the night-side of the planet. The upper part of the atmosphere (67 – 85 km) is more homogeneous and has long-scale horizontal temperature differences of about 25 K over horizontal distances of ~2,000 km. The lower part (55 – 67 km) shows more fine-scale structure, creating the vortex morphology, with thermal differences of up to about 50 K over the same altitude range and ~500 km horizontal distances. This lower part of the atmosphere is highly affected by the upper cloud deck, leading to stronger local temperature variations and larger uncertainties in the retrieval. From the temperature maps, we also study the vertical stability of different atmospheric layers for the three vortex configurations. The static stability is always positive ($S_T > 0$) in the considered altitude range (55 – 85 km) and in the whole polar vortex. The cold collar is the most vertically stable structure at polar latitudes, while the vortex and sub-polar latitudes show lower stability values. Furthermore, the hot filaments present within the vortex exhibit lower stability values than their surroundings. The layer between 62 and 67 km resulted to be the most stable. These results are in good agreement with conclusions from previous radio occultation analyses.

**Keywords**: Venus atmosphere, Atmosphere dynamics, Atmospheres structure, Infrared observations, Radiative transfer

# 1. Introduction

About three decades ago, Mariner 10 ultraviolet images revealed a spiral cloud pattern centered at the north pole of Venus, forming a circumpolar vortex (Suomi and Limaye, 1978). Soon after, the Pioneer Venus infrared data showed the feature as a "double-eye" thermal structure (Taylor et al., 1980; Schofield & Diner, 1983). More recently, the European Venus Express (VEX) mission (Svedhem et al., 2007) confirmed the presence of a similar structure in the south polar region of the planet (Piccioni et al., 2007). On a highly elliptical orbit, VEX monitors the southern hemisphere every 24 hours from apocenter at about 66,000 km above the planet's surface. Both the *Visible and InfraRed Thermal Imaging Spectrometer* (VIRTIS, Drossart et al., 2007) and the *Venus Monitoring Camera* (VMC, Markiewicz et al., 2007) have observed the South Polar Vortex since Venus Orbit Insertion in 2006, showing that it is a permanent feature in the Venusian atmosphere (Piccioni et al., 2007; Luz et al., 2011; Titov et al., 2012; Garate-Lopez et al., 2013).

VMC imagery is obtained in day-light and has a limited visibility of the polar area due to the low solar illumination. A better view of the vortex is obtained with VIRTIS night-side infrared images (1 – 5 µm). These images show that the vortex is confined to latitudes higher than 75ºS (Piccioni et al., 2007) and that it is highly variable in its morphology, evolving through different configurations (dipolar, elongated oval or nearly circular) but being most of the time a transition feature between these configurations (Garate-Lopez et al., 2013). In some cases the vortex preserves an identifiable shape for a few days, but in general it changes its appearance on timescales of 1-2 days (Luz et al., 2011). In addition to variations in shape, the vortex changes much of its characteristic motions. Recent work (Garate-Lopez et al., 2013) has reported that about half of the time the vortex shows a closed circulation pattern that may be concentric with the clouds' morphological center (as expected) but may also be displaced from it. The other half of the time, convergence or divergence circulation patterns, seemingly associated with vertical velocities of up to 0.16 ms$^{-1}$, appear in the vortex. Furthermore, it was found that the vortex is not centered at the south pole but wanders around it in an unpredictable manner. Images of two altitude levels from two observing windows, 1.74 µm sensitive to cloud features at ~42 km above the surface (Barstow et al., 2012), and 3.80 or 5.10 µm, sensitive to the thermal structure and cloud features at ~63 km above the surface (Peralta et al., 2012; Titov et al., 2012; Lee et al., 2012; Haus et al., 2014), indicate that these wandering motions are different at both altitudes, thus showing a vertically curved structure. For the upper altitudes, Luz et al. (2011) found that the vortex's rotation center precesses around the pole with a period of 5 to 10 Earth days but this precession is only a first approximation to more complex motions (Garate-Lopez et al., 2013).

Our current knowledge of Venus' south polar vortex is based mainly on its cloud morphology and motions. In order to interpret its dynamical nature, it is critical to characterize the thermal structure of the vortex, which remains poorly constrained. Piccioni et al. (2007) presented a map of the brightness temperature at 5.05 µm of the southern pole of Venus showing the warm dipolar structure of the vortex (250 K) surrounded by a cold ring (210 K). Thermal maps of the vortex area at different altitudes were obtained by Grassi et al. (2008) from VIRTIS data. These maps showed a smooth distribution of temperature with a maximum difference of about 30 K between the vortex and its surroundings at ~65 km.

Before the VEX arrival, the global thermal structure of the Venusian atmosphere was studied by a variety of techniques with data from different space missions. Amongst others, Seiff et al. (1980) studied thermal fluxes measured by the Pioneer Venus descent probes, Roos-Serote et al. (1995) analyzed infrared spectra in the 4.3 µm $CO_2$ absorption band from the Galileo flyby of Venus and Zasova et al. (1999) investigated Venera-15 spectroscopic data in the $CO_2$ band at 15 µm. These measurements were obtained at a variety of epochs, locations over the surface and were sensitive to different atmospheric altitudes.

The Venus Express mission is well equipped to study the thermal structure of the atmosphere. The Venus Radio (VeRa) Science experiment (Häusler et al., 2006) has obtained thermal profiles of the upper atmosphere from 40 to 100 km using radio occultations (Tellmann et al., 2009). The SPectroscopy for Investigation of Characteristics of the Atmosphere of Venus instrument (SPICAV) has gathered thermal data from 80 to 140 km using stellar and solar occultations (Bertaux et al., 2007a, 2007b). These instruments can characterize the vertical temperatures at a single location at a time but cannot provide the thermal field of a large area at a particular time. The VIRTIS instrument is well suited for that task. Night-side thermal emission of the planet in the 3 – 5 µm range is sensitive to both temperatures and cloud opacities. Grassi et al. (2008) developed a method for obtaining thermal profiles in the Venusian night-side mesosphere from radiances measured by VIRTIS-M (the mapping channel of the instrument). This method has been used to produce moderate resolution thermal maps of the polar area from a single orbit (Grassi et al., 2008) and to obtain average temperature fields as a function of latitude, local time, and pressure (Grassi et al., 2010). Additionally, Migliorini et al. (2012) investigated night-side atmospheric temperatures on both hemispheres using VIRTIS-H (the instrument's high-resolution spectral channel) data. Haus et al. (2013) proposed a radiative transfer model and a multi-window procedure to retrieve information about both temperature profiles and cloud parameters of the atmosphere of Venus. This method was subsequently used to produce statistical global maps of temperature and cloud altitudes from VIRTIS-M. The maps show high temperatures and low clouds at both poles consistent with hot vortices with larger thermal variability than the rest of the planet (Haus et al., 2014).

Instead of a statistical study of temperatures in the polar region we aim here to relate the instantaneous dynamics from the wind field to the particular thermal structure of the vortex on different days. In a recent paper, García Muñoz et al. (2013) investigated the radiative transfer problem of thermal radiation from the Venus night-side between 3 and 5 µm with a purpose-built model of Venus' mesosphere. That work explored the impact of the atmospheric temperature, cloud opacity, and the aerosols' size and chemical composition on the emission spectrum and demonstrated the importance of scattering in the upper cloud and haze layers over Venus' mesosphere. In this work we apply the atmospheric model described by García Muñoz et al. (2013) and a variant of the retrieval algorithm detailed in Grassi et al. (2008) to obtain night-side thermal maps of the Venus south polar region between 55 and 85 km altitudes. These maps are discussed in three different dynamical configurations of the vortex whose dynamics in terms of cloud motions has been previously obtained (Garate-Lopez et al., 2013). We also compare the thermal structure with the retrieved motions and study the imprint of the vortex on the thermal field above the cloud level. In section 2 we describe the selected observations and the thermal retrieval procedure. Thermal and static stability results are presented in section 3. Finally, we discuss the relation between thermal structure and cloud motions in the vortex in section 4.

## 2. Observations and Thermal Retrieval

2.1. Selected observations

VIRTIS is an imaging spectrometer that comprises two subsystems (Drossart et al., 2007), a high resolution spectrometer (VIRTIS-H), and a mapping subsystem (VIRTIS-M) which, in turn, combines separate infrared (VIRTIS-M-IR) and visible (VIRTIS-M-Vis) channels. The infrared channel of the mapping subsystem covers the 1.0 – 5.1 µm range with a spectral sampling of 10 nm and has an instantaneous field of view per pixel of 0.25x0.25 mrad$^2$. When the spacecraft is observing the south pole from apocenter, the spatial resolution in the 256x256 pixel image is about 16 km/pixel. The VIRTIS-M-IR data are of particular interest since they allow us to obtain atmospheric temperatures from the same set of images that have been used to retrieve atmospheric winds.

In a previous work (Garate-Lopez et al., 2013), we used VIRTIS-M-IR images to study the polar vortex at 1.74 µm and 3.80 or 5.10 µm over 23 orbits (1 VEX orbit = 24 hours). The morphological structure and the associated wind field at two altitudes, ~42 and ~63 km above the surface, were extracted from the images using a supervised correlation algorithm (Hueso et al., 2009). For the current work we have selected three of the best cases in that study, which combine high density of wind measurements, different morphologies of the vortex and high-quality spectral data over the 3 – 5 µm range in most of the imaged area. These data sets are: orbit 038 (28 May 2006), when the vortex had a dipolar structure with an inverted-S bright filament surrounded by the cold collar, orbit 310 (24 February 2007) with a nearly circular radiant vortex displaced from the south pole by ~7º in latitude, and orbit 475 (08 August 2007) when the vortex showed a complex structure with a central twisted bright filament.

2.2. Temperature Retrieval

Several $CO_2$ vibrational bands are present in Venus' thermal spectrum at 3 – 5.1 µm (1960 – 3333 cm$^{-1}$). These bands exhibit different strengths, being therefore sensitive to different altitudes. It is possible to infer atmospheric temperature profiles by means of inversion techniques over the measured spectra. In this work, we use an inversion relaxation technique that tries to find a best match between an observed and a modeled spectrum by iteratively correcting an initial guess temperature profile until convergence to a final thermal profile (Hanel et al., 2003). The retrieval algorithm largely follows the methodology established by Grassi et al. (2008). Our algorithm, however, differs from that by Grassi et al. (2008) in the treatment of the clouds, as commented on below.

To produce synthetic spectra we use the forward model described by García Muñoz et al. (2013), where the atmosphere of Venus is assumed to be formed by $CO_2$ (96.5%) and $N_2$ (3.5%), and only mode-2 aerosol particles are considered. The aerosol sizes are assumed to follow a log-normal distribution described by an effective radius of 1.09 µm and an effective variance of 0.037. Refractive indices consistent with a composition of $H_2SO_4$:$H_2O$ in the ratio 84.5:15.5 by mass are adopted. Grassi et al. (2014) and Haus et al. (2014) opt for a prescribed profile of the aerosol number density with a single free parameter scaling that profile. Our model, instead, considers that the aerosol number densities decay with altitude $z$ according to

$$n_{aer}(z) = exp(-(z - Z_{cloud})/H_{aer})/\sigma_{\lambda_*} H_{aer} \quad (1)$$

where $Z_{cloud}$ stands for the cloud top altitude, $H_{aer}$ for the aerosol scale height, and $\sigma_{\lambda_*} = 4.5 \times 10^{-8} \, cm^2$ is a reference value for the aerosol extinction cross section at $\lambda_* = 4 \, \mu m$. This means that at 4 µm a nadir optical thickness of one is reached at $Z_{cloud}$. García Muñoz et al. (2013) investigated the sensitivity of the thermal emission spectrum to the chemical composition and droplet size of the aerosols, to which the reader is referred for further details. For simplicity, we assumed the above droplet size distribution in all the temperature retrievals, although there is evidence suggesting that cloud particles within the vortex may be different in size and/or composition than at lower latitudes (Wilson et al., 2008; Lee et al., 2012).

As the radiative transfer equation solver, we used the package DISORT (Stamnes et al., 1988), which presumes a plane-parallel stratified atmosphere and produces monochromatic multiple-scattering radiances at the top of the atmosphere. Spectra were produced from line-by-line calculations on a fine grid, and were subsequently degraded to the VIRTIS spectral resolution. For efficiency's sake, the $CO_2$ optical properties were tabulated over a range of pressures and temperatures, and invoked during the calls to the forward model. We studied the quality of the modeled spectra when using different fine grids (with spectral sampling of 0.07 cm$^{-1}$, 0.13 cm$^{-1}$ and 0.26 cm$^{-1}$), and found that the spectra (after degrading them to the VIRTIS resolution) did not show great differences. Thus, to reduce the computational cost we used the 0.26 cm$^{-1}$ resolution sampling, resulting in more than 1500 monochromatic calculations per spectrum in the spectral range of interest (1965 – 2380 cm$^{-1}$).

We initialized the retrieval with an initial guess thermal profile based on VeRa measurements at around 70ºS (Lee et al., 2012), smoothed to the vertical resolution of our retrieval, thus avoiding small wiggles present in the guess profile. Then we iteratively improved the solution profile using the relaxation algorithm given by Grassi et al. (2008) in their equation (2), which describes the temperature update between two consecutive iterative cycles for each pressure level. In our case, the atmosphere between 55 and 85 km altitude was modeled in 30 pressure levels, with special care being paid to the lowermost layers. We took into account the exact value of the emission angle, $e$, which can be as high as $e \sim 77°$ in one of the orbits analyzed (orbit 038) and presents large variations of up to about 63° throughout the image. Tests made during the validation process showed that omitting $e$ may lead to differences in the retrieved temperatures larger than the estimated error (discussed below).

The algorithm does not fit simultaneously well both shoulders of the wide 4.3 μm CO$_2$ absorption band. Similar difficulties have also been noted by Roos-Serote et al. (1995) and Grassi et al. (2008). The sensitivity study presented in García Muñoz et al. (2013) shows that the spectral region from 4.2 to 5.1 μm may contain enough information to retrieve the temperature from about 55 to 85 km, especially for low cloud altitudes. However, the weighting functions associated to the 4.5 - 5.0 μm wavelengths (which sense the deepest altitudes considered in this work) are notably wider than those associated to higher altitudes, meaning that separation of contribution from different altitudes is more complicated and that the uncertainty increases rapidly in the lowest levels, being particularly large between 55 and 60 km. We decided to fit the long-wavelength shoulder of the 4.3 μm band, between 4.2 and 5.1 μm, but excluding the spectral region dominated by the CO 1-0 fundamental band from 4.52 to 4.80 μm (2083 – 2212 cm$^{-1}$) (Grassi et al., 2008).

The goodness of the fit was measured by calculating the root-mean-square deviation (RMSD) of the brightness temperature of the retrieval compared with the observed brightness temperature (the brightness temperature at a given wavelength is the temperature of a black body with the same emitted radiance):

$$\chi = \sqrt{\frac{\sum_{i=1}^{n}(TB_i^{obs}-TB_i^{mod})^2}{n}} \qquad (2)$$

where each *i* represents one of the 80 VIRTIS sampling channels between 1965 and 2380 cm$^{-1}$, except those falling in the region dominated by the CO band.

We found that the observed and modeled spectra were similar enough (being the mean value $\chi \sim 1.35\,K$) after 20 steps in the iteration process for every retrieval we tried. In most cases, after the third iteration cycle the RMSD value decreased by ~80% (this translates into a difference of $\Delta\chi \sim 1K$ between the second and third steps) and continued decreasing in the following iterations. To ensure full convergence in all cases, we systematically ran the algorithm over 20 iterations. Lower values of $\chi$ were not achievable by increasing the number of iterations, which is related to the intrinsic noise in the spectrum and the model characteristics. In particular, the instrument Noise Equivalent Radiance (NER) at 4.3 µm is 5x10$^{-4}$ Wm$^{-2}$ster$^{-1}$µm$^{-1}$ (Grassi et al. 2008). Since we averaged the spectra in 2x2pixel bins, the instrumental noise is reduced by a factor 2. The goodness of the fit is 3.5x10$^{-4}$ Wm$^{-2}$ster$^{-1}$µm$^{-1}$ as measured by means of the radiance difference between observed and modeled spectra at 4.3 µm and is comparable to the NER. The goodness of the fit is quite homogeneous over the complete night-time polar area and there is no region where the fit fails to converge in any of the three orbits analyzed. The algorithm proved to be quite robust, always producing reliable fits over the 4.2 – 5.1 µm region.

The retrieved temperatures are generally consistent with what is known about the Venus atmosphere. For example, some individual temperature profiles show thermal inversions at locations where the cold collar exists (65 – 70ºS, 100 mbar) (Taylor et al., 1980; Tellmann et al., 2009), and at high altitudes the polar regions were warmer than at lower latitudes. Importantly, the retrieval procedure is "fast" enough (an individual retrieval takes about 2.3 minutes on a Xeon 2.5GHz computer when using one of its cores), thus, enabling the analysis of a complete VIRTIS cube in ~12 days in a multi-core server and making it possible to construct temperature maps with very high spatial resolution. Figure 1 shows examples of the retrieval for individual locations over the vortex and the surrounding cold collar for the VI0038_00 data cube.

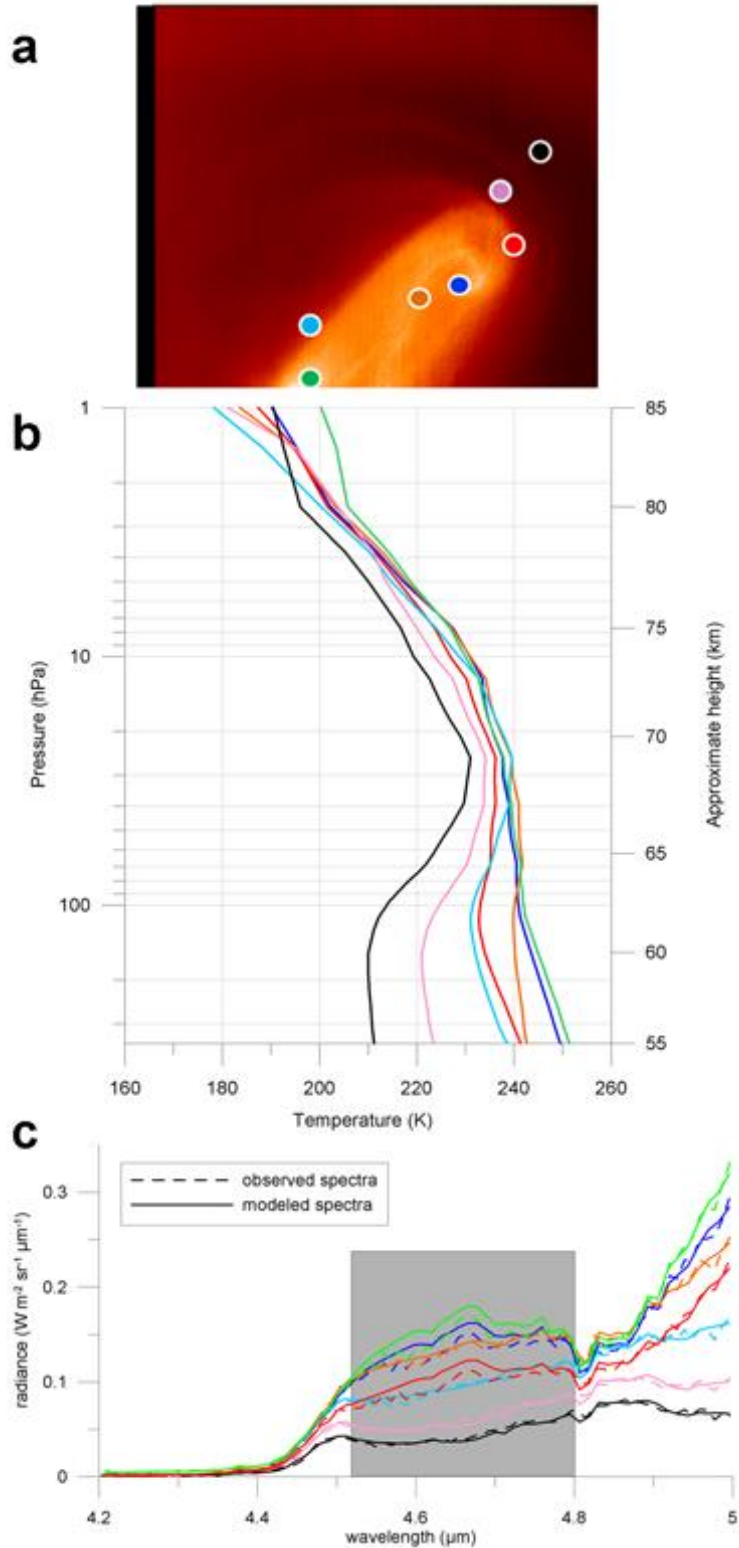

**Figure 1:** (a) 5 μm image of the polar vortex obtained on orbit 038. Seven individual positions over the vortex and its surroundings are indicated. (b) Retrieved thermal profiles for the corresponding color points in panel a. (c) Comparison of the modeled and observed spectra. The shaded region from 4.52 to 4.80 μm is excluded in the algorithm, which explains the differences between both sets of spectra in that region.

2.3. Cloud parameters

The cloud top altitude ($Z_{cloud}$) and the aerosol scale height ($H_{aer}$) have a notable impact on the modeled spectra. As noted by Grassi et al. (2008), it may indeed be difficult to discriminate between the effects of these two parameters. Our approach to the treatment of clouds in the retrieval differs from that of Grassi et al. (2008) because we do not consider $H_{aer}$ and $Z_{cloud}$ as free parameters within the iteration algorithm. Instead, we retrieve temperatures for fixed values of $H_{aer}$ (2 and 4 km) and $Z_{cloud}$ (56, 58, 60, 62, and 64 km), thus sampling the space of $H_{aer}$ and $Z_{cloud}$ parameters. Our educated choice of $H_{aer}$ and $Z_{cloud}$ was based on a number of experiments previously run with a larger set of combinations. The values of $H_{aer}$ and $Z_{cloud}$ that generally showed a better fit of the modeled and observed spectra (see equation 2 above) were selected to be used in the temperature retrievals of the three orbits under investigation.

In order to reduce the computational time, the retrieval was divided into two steps, each of them considering a different spatial resolution. The first step works with spectra averaged over 6x6 pixels, and it serves to infer the $H_{aer}$ and $Z_{cloud}$ values that minimize the spectrum averaged over that 6x6 pixel domain. In this step, we proceeded to run the retrieval with the 2x5 combinations of $H_{aer}$ and $Z_{cloud}$ listed above, obtaining ten different thermal profiles. Then, we compared the modeled spectra with the average spectrum and identified the spectrum and cloud parameter combination resulting in the smallest $\chi$ residual. The second step works with spectra averaged over 2x2 pixels. At this finer spatial scale, we proceeded to retrieve the temperature with the values of $H_{aer}$ and $Z_{cloud}$ obtained in the minimization of the average spectrum of the 6x6 boxes. Hence, our final 2x2 pixel temperature maps have half the spatial (horizontal) resolution than that in the original VIRTIS-M-IR images, 30 – 34 km/bin, and our 6x6 pixel maps of cloud parameters a spatial resolution of ~100 km/bin.

Although our method is not well suited for the retrieval of the cloud properties, it is interesting to note that the spectra corresponding to the cold collar are generally better fit when using a lower value of the aerosol scale height, while the vortex and the sub-polar latitude atmosphere seem to be better reproduced with a larger $H_{aer}$ (see Figure 2). Lee et al. (2012) and Zasova et al. (1993, 2007) reported a decrease in $H_{aer}$ towards the poles, but they did not analyze the horizontal structure within the polar region.

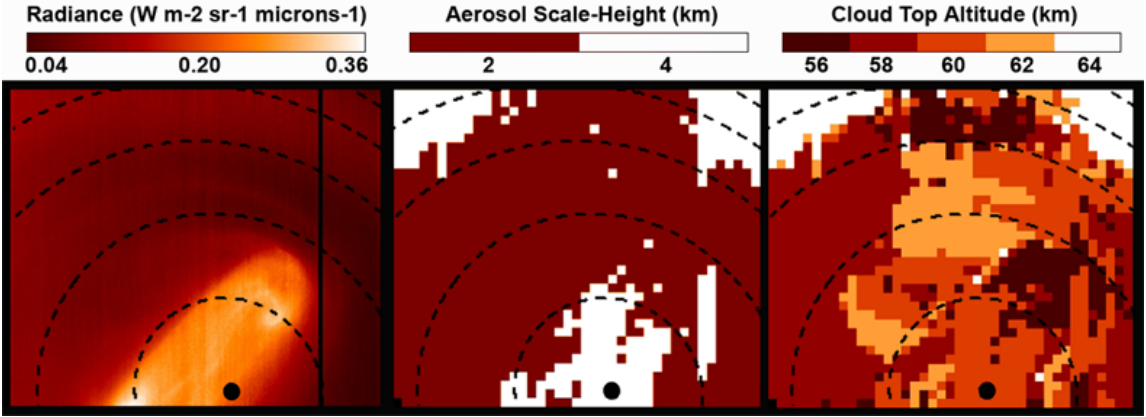

**Figure 2:** (a) Vortex morphology as seen in radiance images at 5 μm for the VI0038_00 data cube. (b) $H_{aer}$ distribution and (c) $Z_{cloud}$ distribution in the 6x6 pixel boxes, in terms of "optimal fit" between measured and modeled spectra defined by equation (2). Black dashed lines depict isolatitude lines each 10º from the south pole.

Latitudes lower than those associated with the cold collar seem to be better fit with a higher value of the cloud top altitude (about 4 km higher than in the cold collar or vortex), which is in agreement with the idea of the polar region being a depression area (Titov et al., 2008; Ignatiev et al., 2009). Recently, analyzing VIRTIS-M-IR night-side spectra for the southern hemisphere, Haus et al. (2014) reported that the altitude of the cloud top varies by about two atmospheric scale heights over the planet, being lower close to the pole. Furthermore, they found an altitude decrease of ~6 km from 60 to 80ºS (see fig. 17 by Haus et al., 2014). Ignatiev et al. (2009) also detected a decrease of ~7 km from 60ºS to the pole and found a correlation of the fine structure of the vortex eye with the cloud top pattern. Nevertheless, our variability of the $Z_{cloud}$ values (and to a lesser extent of the $H_{aer}$ values) that result in better spectral fits within the cold collar and vortex area is high and prevents us from setting constraints on the spatial structure of the cloud parameters. Hence, the inferred values of $H_{aer}$ and $Z_{cloud}$ should be taken only as indicative.

A comparison of cloud altimetry from different works needs to be done with caution (Lee et al., 2012). $Z_{cloud}$ is defined as the altitude where the cumulative cloud optical depth as measured from the top of the atmosphere becomes unity: $\tau_{nadir} = \int_{Z_{cloud}}^{\infty} \sigma_{\lambda_*} n_{aer}(z)dz = 1$. Thus, $Z_{cloud}$ is wavelength dependent. In the present work we used $\lambda_* = 4\ \mu m$, which means that our cloud top altitude levels are about 4 km lower than those reported by Ignatiev et al. (2009) using the $CO_2$ band in the visible range (see Figure 11 below for a comparison of the cloud top altitude derived from different observations).

## 2.4. Sensitivity analysis and temperature errors

We studied the accuracy of the retrieval with an exercise based on synthetic spectra. For the exercise, six thermal profiles obtained from the VeRa experiment (Y. J. Lee, personal communication) and representative of different latitudes over the planet were smoothed to our vertical sampling of 30 pressure levels and used to build 40 synthetic spectra. These synthetic spectra, corresponding to "true temperatures" and modeled clouds, were used to perform a sensitivity analysis of the retrieval.

For each synthetic spectrum we retrieved 2x5 thermal profiles corresponding to the $Z_{cloud}$ (56, 58, 60, 62, and 64 km) and $H_{aer}$ (2 and 4 km) utilized in the usual retrieval process. Figure 3 displays one example showing the "input" VeRa profile (measured at polar latitudes) and the retrieved profile that minimizes the $\chi$ residual. The thermal profile used as an initial guess in these tests and in the usual retrieval exercises is also shown.

Error bars representative of the systematic sensitivity analyses for the 40 cases studied are also shown in Figure 3, and call for an in-detail description. We define two types of errors: the retrieval error (which contains the errors associated with temperature and cloud structure) and the algorithm error (which only contains the errors associated with the retrieval of the temperatures and are computed from cases where the assumed cloud structure is known). Both errors represent the average of the differences between the retrieved temperatures and the true temperatures of the VeRa profile. In the retrieval error (dashed error bars in Figure 3) the selected retrieved profile is the one associated with the best-fit case for each synthetic spectrum (the $H_{aer}$ x $Z_{cloud}$ combination that results in the smallest $\chi$ residual). In the algorithm error (labels and solid errors in Figure 3) the selected retrieved profile is that with a cloud model equal to the one used to generate the synthetic spectra.

In view of Figure 3, the algorithm uncertainties are about 3 K on average between 360 mbar (~55 km) and 1 mbar (~85 km), but increase rapidly in the lowest ~7 km where they can be as large as 9 K. This is related to the difficulty of probing levels below 60 km using VIRTIS data in the 4.2 – 5.1 µm spectral range. Moreover, the uncertainty related to the lack of knowledge of the true cloud structure becomes significant below 60 km. However, and since we utilized $Z_{cloud}$ values as low as 56 km, we retrieved temperature profiles down to the 55 km although we note that these deeper levels are much more affected by uncertainties. At altitudes above the 100 mbar level, our errors are slightly larger than those reported by Grassi et al. (2008). Below that level, the retrieval errors considering the clouds effects are similar to those reported by Grassi et al. (2008).

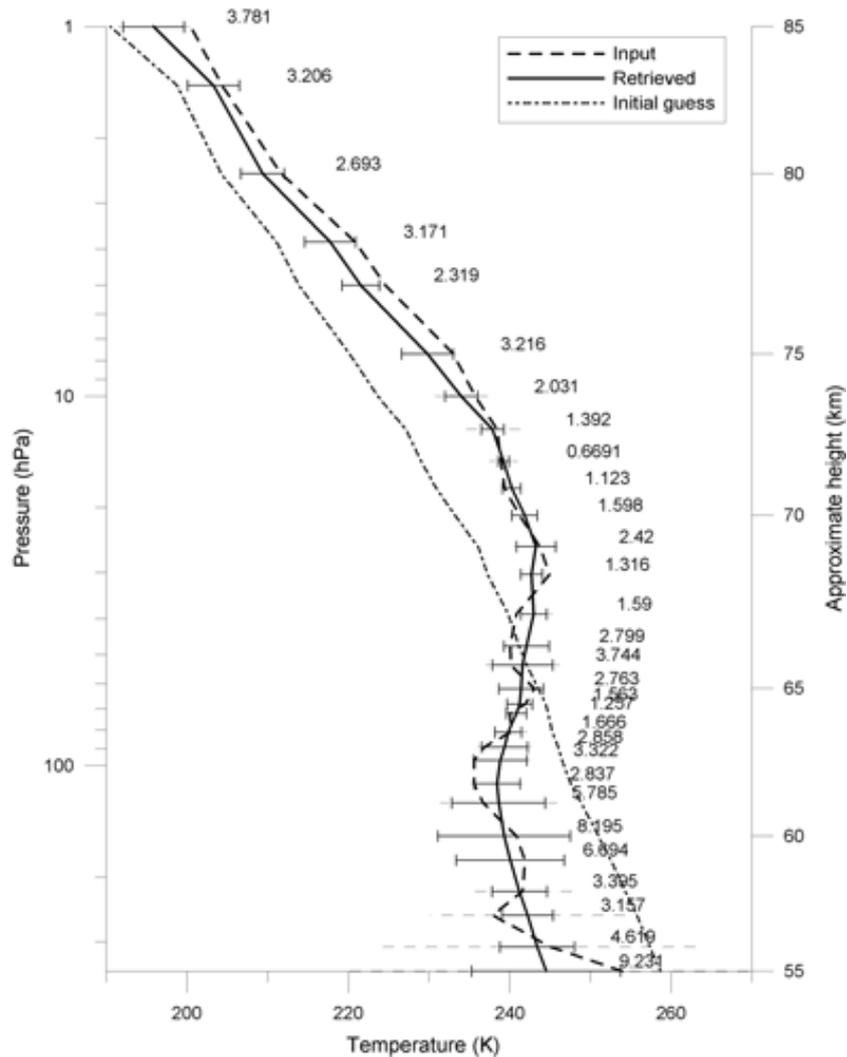

**Figure 3:** Example of retrieval and sensitivity analysis. An input thermal profile from a true VeRa thermal profile representative of high latitudes (dashed line), together with a cloud model is used to produce a synthetic spectrum which is the starting point for a thermal retrieval. Starting from our initial guess (dotted line) the retrieval produces a final thermal profile (continuous line) and values that characterize the cloud model ($Z_{cloud}$ and $H_{aer}$). Error bars and numbers represent statistical errors for several retrievals and are estimated as explained in the text. Dashed error bars represent the retrieval error including the uncertainty in cloud parameters. Solid error bars and labels show the algorithm error when the cloud parameters of the retrieval are fixed to the values considered in the synthetic spectrum.

## 3. South Polar Vortex Results

3.1. Temperature Structure

In the Venusian atmosphere the temperature increases downwards from ~100 km to ~40 km, except in an inversion layer (at about 60 - 70 km) coincident with the cold collar (Taylor et al., 1980; Seiff et al., 1983; Piccialli et al., 2008; Tellmann et al., 2009). In general, our retrieved thermal profiles agree well with the expected trend mentioned above and reproduce the inversion at the corresponding latitudes (see panel b in Figure 1). Due to the vertically extended contribution functions of thermal radiation, our retrievals are smooth functions, which likely cause a smoothing of the inversion layer. Other studies such as Zasova et al. (2007), Tellmann et al. (2009), and Lee et al. (2012) are more sensitive to temperature changes over small vertical distances. Our profiles are more similar to those obtained by Haus et al. (2014) from VIRTIS data.

Figures 4, 5, and 6 show temperature maps for different altitude levels retrieved from data cubes VI0038_00, VI0310_00, and VI0475_04, respectively. These cubes fulfill the conditions described in section 2.1 (high density of wind measurements, different morphologies of the vortex and high-quality spectral data over the 3 – 5 µm range in most of the imaged area). In the three orbits under investigation the middle mesosphere (panel a; pressure level of 1 mbar and altitude of ~85 km) does not show any thermal feature while the structure of the vortex begins to emerge at pressures of 5 mbar (shown in panel c). So, it seems that the thermal signature of the South Polar Vortex on the night-side goes from at least the upper clouds up to ~80 km (~3 mbar).

As we go downwards in the atmosphere from 5 to 35 mbar (i.e. from ~78 to 68 km), the maps show the structure of a warm region surrounded by colder air. At 90 mbar (~63 km) the "hot vortex" and the "cold collar" are clearly discernible. In the two higher pressure levels, 155 mbar (~60 km) and 360 mbar (~55 km), the vortex shows detailed thermal structure at small scales with narrow hot filaments while the cold collar remains as an horizontally-homogeneous region. In the lower atmospheric levels the retrieved temperatures are more sensitive to the clouds parameters and multiple scattering, which causes additional uncertainties. The noise seen in these maps is affected by the sensitivity to the cloud parameters. The temperature distribution at 360 mbar, although probably uncertain by ~9 K (as discussed earlier), compares very well with the radiance image at ~5 μm for the three vortex configurations of our study.

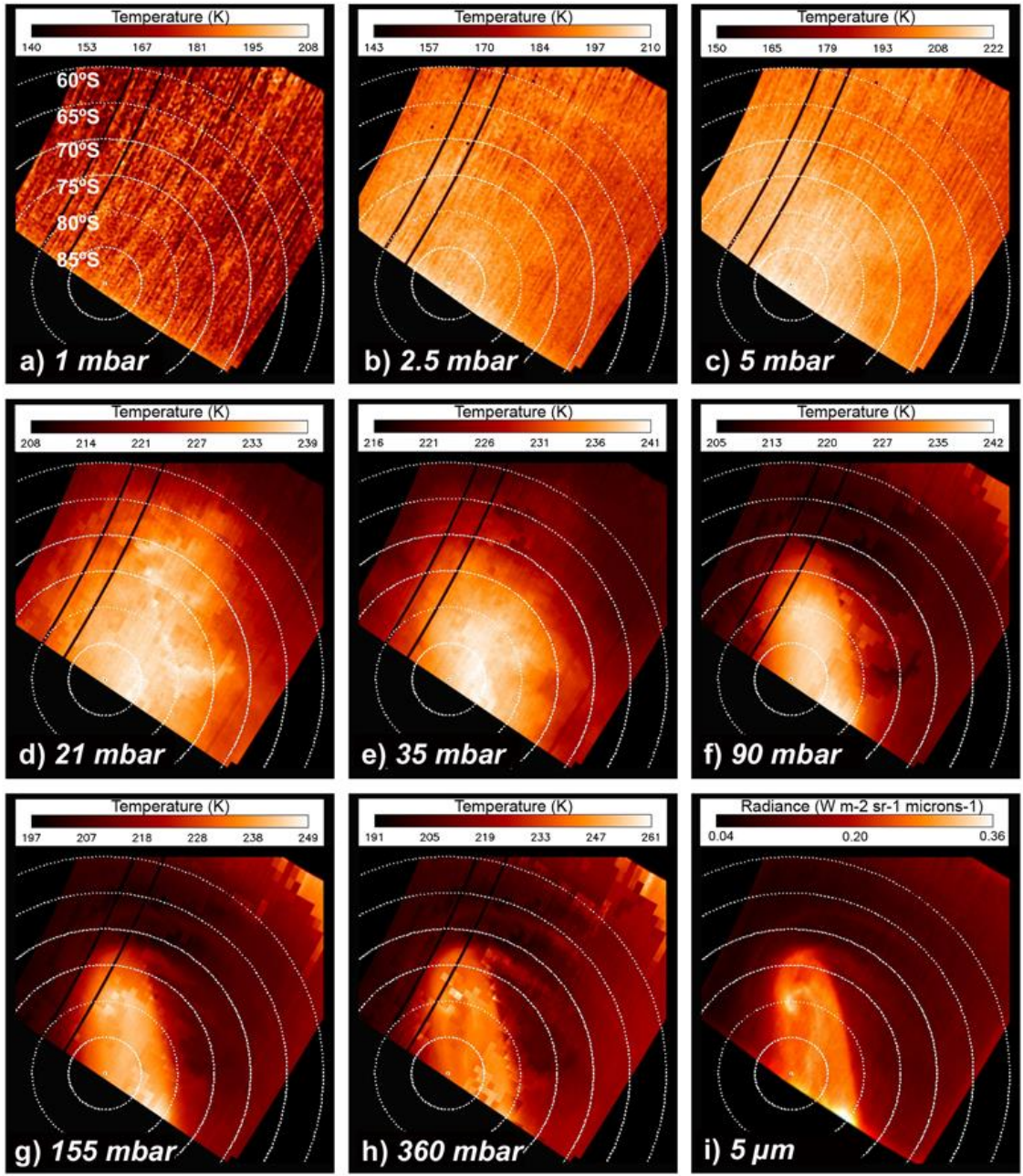

**Figure 4:** Retrieved temperature maps at atmospheric pressure levels of (a) 1 mbar (~85 km), (b) 2.5 mbar (~81 km), (c) 5 mbar (~78 km), (d) 21 mbar (~70 km), (e) 35 mbar (~68 km), (f) 90 mbar (~63 km), (g) 155 mbar (~60 km), and (h) 360 mbar (~55 km), and radiance image over the south polar region of Venus as measured on orbit 038 at ~5 μm (i).

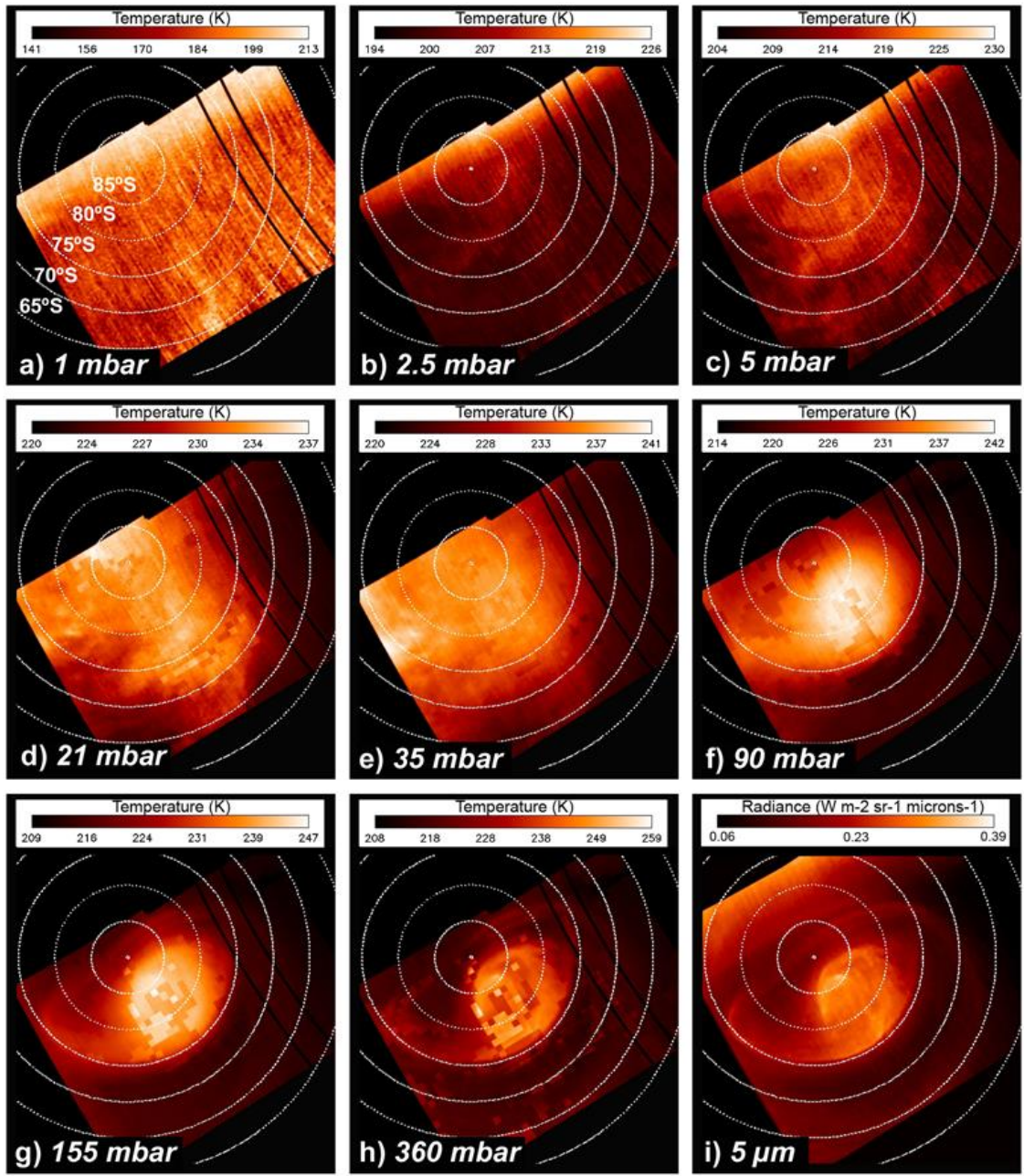

**Figure 5:** Retrieved temperature maps at atmospheric pressure levels of (a) 1 mbar (~85 km), (b) 2.5 mbar (~81 km), (c) 5 mbar (~78 km), (d) 21 mbar (~70 km), (e) 35 mbar (~68 km), (f) 90 mbar (~63 km), (g) 155 mbar (~60 km), and (h) 360 mbar (~55 km), and radiance image over the south polar region of Venus as measured on orbit 310 at ~5 µm (i).

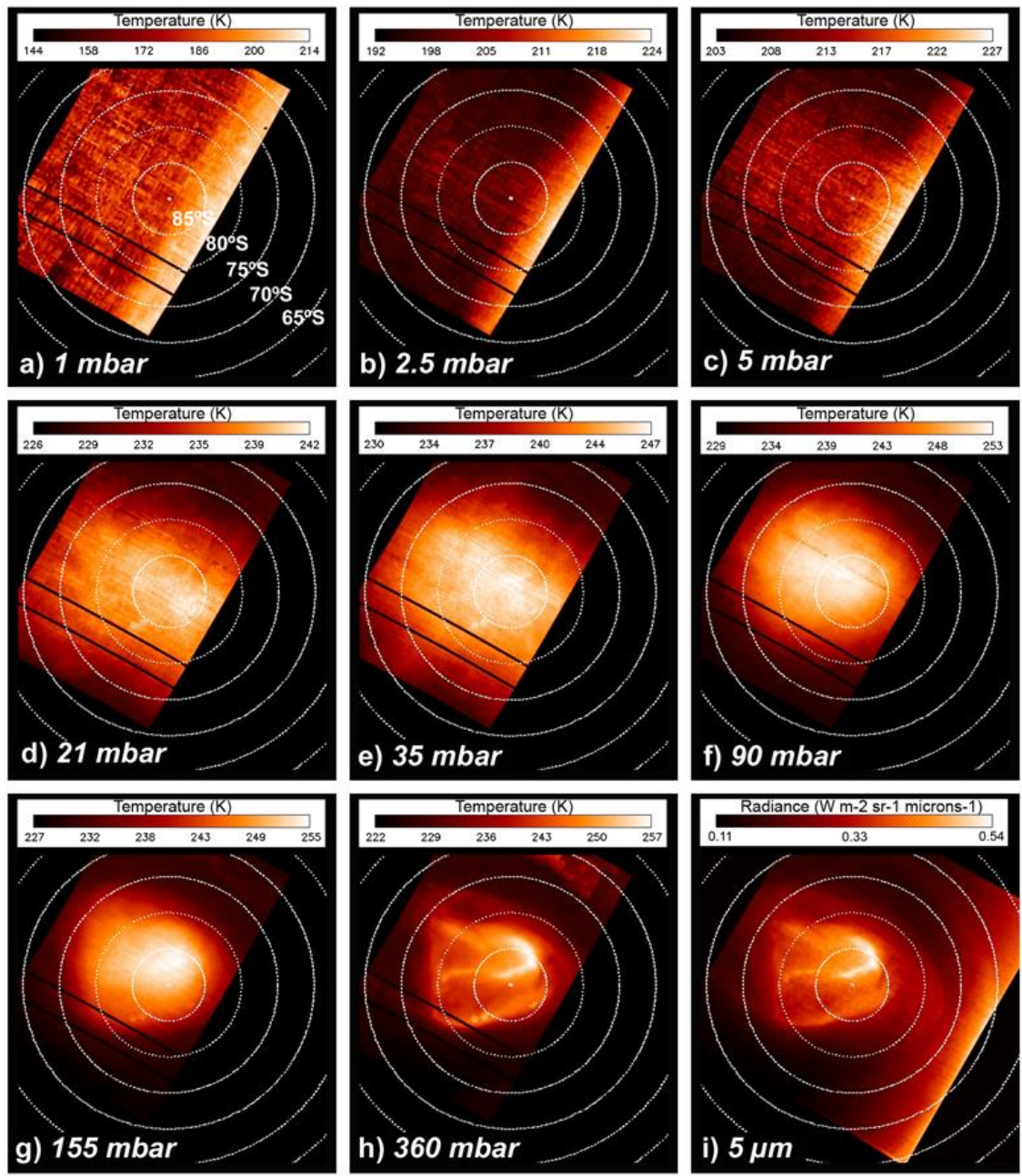

**Figure 6:** Retrieved temperature maps at atmospheric pressure levels of (a) 1 mbar (~85 km), (b) 2.5 mbar (~81 km), (c) 5 mbar (~78 km), (d) 21 mbar (~70 km), (e) 35 mbar (~68 km), (f) 90 mbar (~63 km), (g) 155 mbar (~60 km), and (h) 360 mbar (~55 km), and radiance image over the south polar region of Venus as measured on orbit 475 at ~5 µm (i).

Temperature differences between the "hot vortex" and the "cold collar" are larger than the associated retrieval errors at every altitude level. At the 35 mbar level, the collar is on average ~13 K colder than the mean vortex temperature and this difference increases downwards. At 360 mbar, it is ~30 K on average, but the particular value varies from orbit to orbit (as seen on panel h of Figures 4, 5, and 6). These results are overall consistent with the temperature differences found by Piccialli et al. (2008) and Tellmann et al. (2009), and agree also with the maximum absolute temperature variations of about 35 K at 65 km and high latitudes (70 – 80ºS) given by Haus et al. (2014). In addition, Zasova et al. (2007) suggested that the temperature differences between the hot vortex and cold collar near the upper boundary of the clouds may exceed 50 K. We found that the maximum horizontal temperature gradient is given between the hot vortex and the cold collar at the pressure level of 155 mbar on orbit 038. The estimated value is of the order of dT/dr = 0.1 K/km (being r the horizontal distance), which translates into a maximum of ~50 K of difference over a distance of about 500 km between the hot vortex and the cold collar.

Even if the temperature differences in the lowest two levels are below the algorithm's estimated error (~3 K at 90 mbar or ~9 K at 360 mbar, see Figure 3), there is an apparent resemblance between temperatures and radiances at ~5 µm, which suggests that the temperature differences are genuine. Between ~55 and ~60 km, the temperature gradient from the hot vortex to the cold collar and the local thermal variations within the vortex or the cold collar are larger than in the upper altitudes.

On orbit 038 (Figure 4), the vortex shows the well-known dipolar shape. It also shows the largest variations in temperature for each pressure level when compared with the different morphologies analyzed in this work. Our temperature maps at 35 and 90 mbar show more structure than the temperature fields obtained by Grassi et al. (2008) due to the higher resolution used in our retrieval, but we note that the general behavior of the two works is largely consistent. On the top-right corner of panels f-g-h (Figure 4) there is a hotter region that defines the end of the cold collar at about 55ºS. On orbit 310 (Figure 5) the vortex is more circular and can be seen completely on the night-side, being displaced from the south pole by ~7º in latitude. It seems that the 75ºS latitude circle divides the polar region in two. The higher latitudes show the "hot vortex", while the cold collar is located between 75 and 65ºS. The missing section of the image was omitted from the retrieval for being on the day-side.

On orbit 475 (Figure 6), the vortex has an irregular shape characterized by a winding warm filament that is only seen on the thermal maps at the lowest altitude level (~55 km). In this case, the optimal fit for the cloud parameters is very homogeneous within the vortex (low $H_{aer}$ and $Z_{cloud}$ values), resulting in clear structures visible in the thermal maps at the lowermost altitudes.

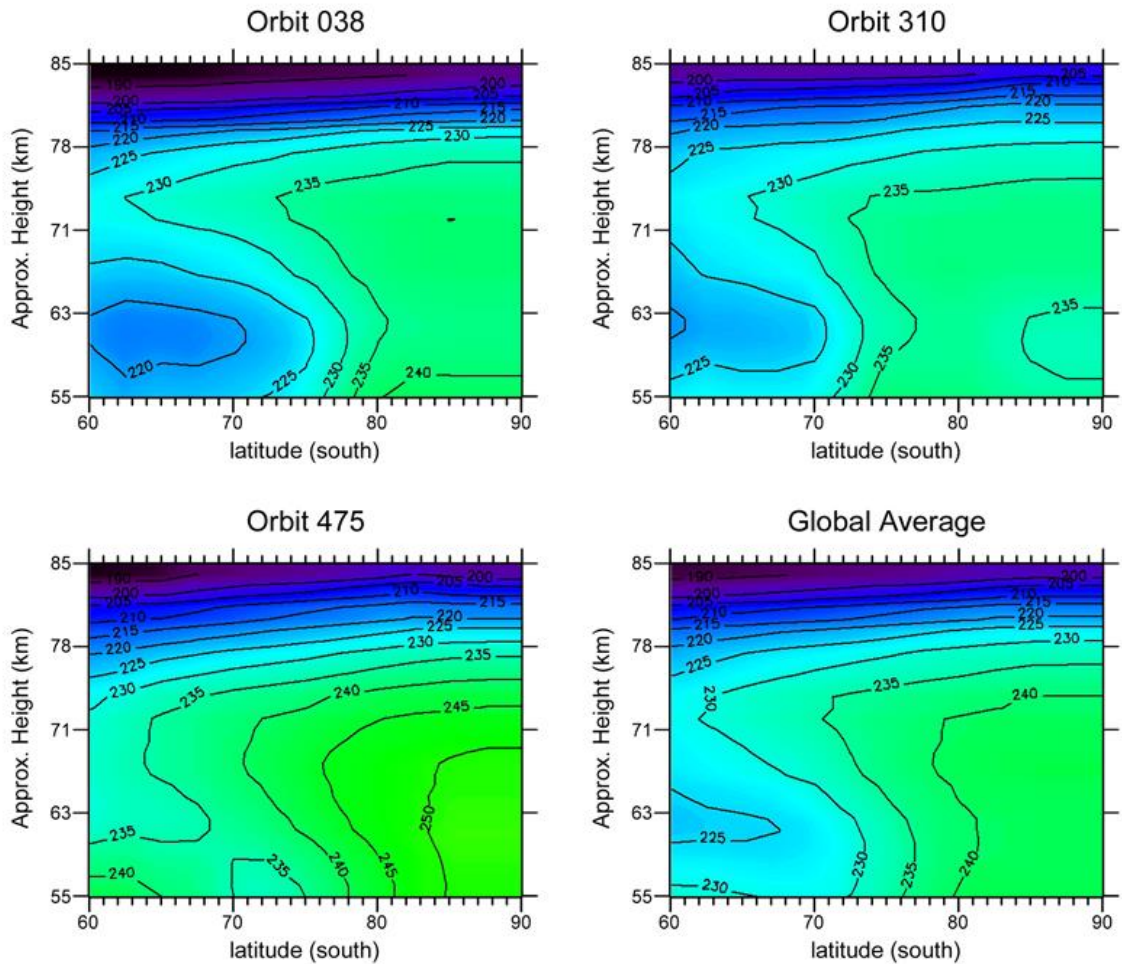

**Figure 7:** Zonal average thermal maps on orbits 038, 310, and 475, and mean temperature field between altitudes 55 and 85 km.

Figure 7 shows zonally averaged thermal maps as functions of altitude for each orbit and the retrieved atmospheric mean temperature field. The night-side thermal distribution is similar for the three cases, with a distinct cold collar centered at an altitude of ~62 km and at latitudes lower than 75ºS. On average, it is more than 15 K colder than the pole, but its specific temperature varies from orbit to orbit. Temperature values for the cold collar are in good agreement with those obtained by Tellmann et al. (2009), Lee et al. (2012) and Haus et al.

(2014). The vertically extended hot region close to the pole represents the vortex, squeezed by the cold collar between 55 km and ~67 km but spreading equatorwards at about 74 km. The top of the thermal signature associated with the South Polar Vortex is clearly seen at ~78 km, where the upper atmosphere shows a mild increase of temperatures towards the pole. This agrees well with the fact that the cold collar region divides the atmosphere vertically (Tellmann et al., 2009). Above the collar, the atmosphere becomes warmer with increasing latitude, while below it the temperature gradient is reversed. Unfortunately VIRTIS spectra between 4.2 and 5.1 µm do not probe altitudes lower than ~55 km, which would offer additional insight into the roots of the vortex.

If we compare the three study cases, on orbit 038 the cold collar is more pronounced showing temperatures on the order of 220 K, while on orbit 475 the cold collar temperatures increase to 235 K. The latter case also shows the highest temperature values for the hot vortex. Orbit 310 exhibits a slightly colder region centered at the pole and at altitude ~60 km, probably due to the vortex being displaced from the south pole, as seen in Figure 5.

Due to the small number of orbits analyzed, it is difficult to assess whether differences in temperature and cloud parameters from orbit to orbit are part of the dynamics of the vortex or if they rather represent outliers in the general behavior of the vortex. However, since the temperature maps at 360 mbar correlate well with the images at 3.8 and 5.1 µm and the cases studied here exhibit three characteristic shapes of the vortex (while orbit 38 represents the classical "dipolar vortex", orbits 310 and 475 are good examples of the "single eyed" vortex and the transition complex features found in many orbits (Garate-Lopez et al., 2013)), we believe that they represent well the various possible thermal configurations of the vortex.

3.2. Static Stability maps

From the temperature maps, we also studied the vertical stability of different atmospheric layers on the night-side of the planet. For that purpose, we calculated the static stability, $S_T = dT/dZ + g/C_P$, where $\Gamma = g/C_P$ is the adiabatic lapse rate of the atmosphere and $g$ and $C_P$ are the gravity acceleration and specific heat of the atmosphere (Sánchez-Lavega, 2011). The static stability measures the gravitational resistance of an atmosphere to vertical displacements, and is generally given in K/km, so that the atmosphere is stable as long as $S_T > 0$.

The specific heat of a $CO_2$ atmosphere depends on temperature. We estimated $C_P(T)$ with the approximation described in Epele et al. (2007),

$$C_P/R^* \sim A + BT + CT^2, \qquad (3)$$

where $R^* = R/M_{CO_2}$ ($R = 8.3143\ J\ mol^{-1}K^{-1}$ and $M_{CO_2} = 44.01\ g\ mol^{-1}$) and the coefficients *A*, *B* and *C* have been empirically adjusted resulting in $A = 2.5223$, $B = 0.77101 \times 10^{-2}\ K^{-1}$, and $C = -0.3981 \times 10^{-5}\ K^{-2}$.

Since $S_T$ depends on vertical derivatives of temperature it is necessary to use an adequate vertical discretization that minimizes errors in the derivative while preserving the information on $S_T$. Therefore, we considered relatively thick vertical layers in order to maintain the estimated static stability errors below a ~10% of the adiabatic lapse rate (Γ ~10.4 K/km on the Venus atmosphere). The altitude range ($\Delta Z$) for each atmospheric layer where the static stability errors fulfill this constraint was calculated based on the average value of the differences in vertical temperature gradient of a known input profile and the retrieved profile:

$$\delta S_T = \left\langle \left| \left(\frac{\Delta T}{\Delta Z}\right)_{input} - \left(\frac{\Delta T}{\Delta Z}\right)_{retrieved} \right| \right\rangle \qquad (4)$$

where "input" refers to the same VeRa thermal profiles used in the retrieval's error estimation. In this way, the atmosphere from 55 km to 85 km was divided in seven layers (the upper one not shown here) for which $S_T$ errors remain lower than 1 K/km except in the deepest layer where it is 1.70 K/km (~16.4% of Γ). Despite these moderate uncertainties, the local differences are larger than the errors and the horizontal structure of the static stability retains the general shape of the vortex, thus confirming the obtained results. The thickness for each layer from top to bottom is equal to 4.0, 3.2, 3.3, 3.7, 3.2, 4.9, and 7.3 km.

In the three cases under analysis, the atmosphere from ~55 km up to ~85 km is highly stable with two differentiated behaviors (see Figures 8, 9, and 10). Above ~74 km the temperature does hardly show any structure, and the static stability is about 6 – 7 K/km everywhere, which is in agreement with the VeRa data presented by Tellmann et al. (2009). In the lower two layers (from ~55 km to ~67 km), the cold collar appears as the most stable feature and the sub-polar latitudes present the lowest values of $S_T$. The hot filaments present within the vortex exhibit lower stability values than the overall vortex. The stability difference between the cold collar and the vortex is not surprising since the cold collar is a permanent feature in the Venusian atmosphere while the vortex evolves constantly. The intermediate layers between ~67 km and ~74 km are more variable, showing mixed regions of higher/lower stability (orbits 038 and 310), or being almost flat as it is the case in orbit 475. It is difficult to know whether the different behavior of orbit 475 represents a statistically significant case or not.

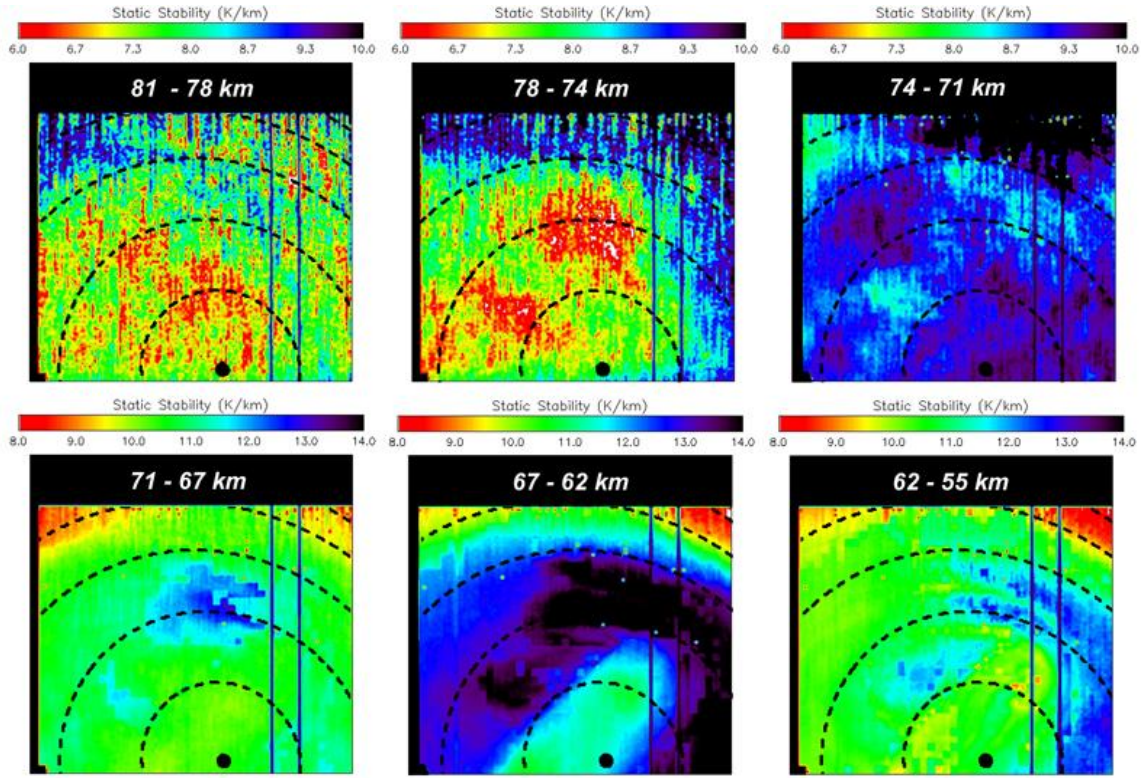

**Figure 8:** Static stability distribution on orbit 038 for several atmospheric layers (a) 81 – 78 km, (b) 78 – 74 km, (c) 74 – 71 km, (d) 71 – 67 km, (e) 67 – 62 km, and (f) 62 – 55 km. The estimated errors are 0.5 K/km, 0.4 K/km, 1.0 K/km, 0.5 K/km, 0.6 K/km and 1.7 K/km, respectively. Black dashed lines depict isolatitude lines each 10º from the south pole. The cold collar is the most stable feature between 55 and 67 km. Note the scale change between the top and bottom rows.

The thin atmospheric layer between 62 and 67 km, which is located just above the cloud top, is the region where maximum values of the static stability are found. This coincides with the highly stable layer centered at 64 km found by Tellmann et al. (2009). At this layer, the static stability reaches values as high as 14 K/km at the cold collar, while in sub-polar latitudes it is lower than 10 K/km (Figure 8). The stability difference between the vortex and the cold collar on orbit 038 is about 4 K/km, and the general oval shape of the vortex is clearly distinguished as a feature of lower stability surrounded by the highly stable cold collar. In the second orbit of our analysis, Figure 9, the stability within the vortex remains on the order of 10 K/km, but the surrounding areas show slightly lower values than in the previous case. On orbit 475, the entire polar atmosphere has lower stability values, but the difference of ~2 K/km between the vortex and the cold collar remains.

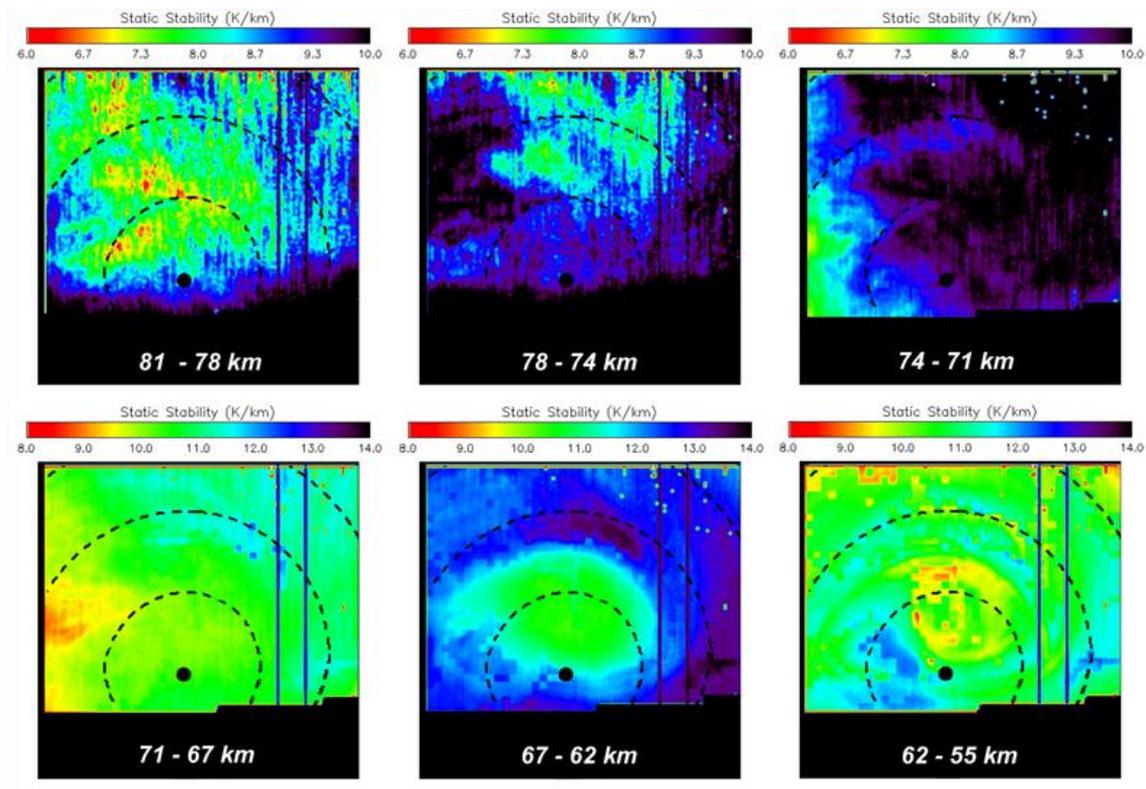

**Figure 9:** Static stability distribution on orbit 310 for several atmospheric layers (a) 81 – 78 km, (b) 78 – 74 km, (c) 74 – 71 km, (d) 71 – 67 km, (e) 67 – 62 km, and (f) 62 – 55 km. The estimated errors are 0.5 K/km, 0.4 K/km, 1.0 K/km, 0.5 K/km, 0.6 K/km and 1.7 K/km, respectively. Black dashed lines depict isolatitude lines each 10º from the south pole. The cold collar is the most stable feature between 55 and 67 km. In this altitude range the hot filaments in the vortex are the less stable features. Note the scale change between the top and bottom rows.

In the lowest layer (between 55 and 62 km) the static stability is smaller and seems to decrease as we go down in the atmosphere (mildly seen in the lower layers of Figures 8 and 9). Unfortunately, our temperature retrieval method cannot estimate thermal information below 55 km, and the associated uncertainty grows in the lower atmosphere. Consequently, derived quantities such as the static stability become quickly uncertain, showing relatively large errors between 55 and 62 km. Importantly, despite the error of about 1.7 K/km, the overall shape of the vortex is retained. So, even if the middle cloud deck (about 48 – 58 km) is potentially unstable, and shallow convective regions may exist there (Tellmann et al., 2009), the VIRTIS spectral range does not allow us to analyze $S_T$ in detail at the corresponding altitude levels, thus preventing the study of the relation between these convective regions and the vortex variability.

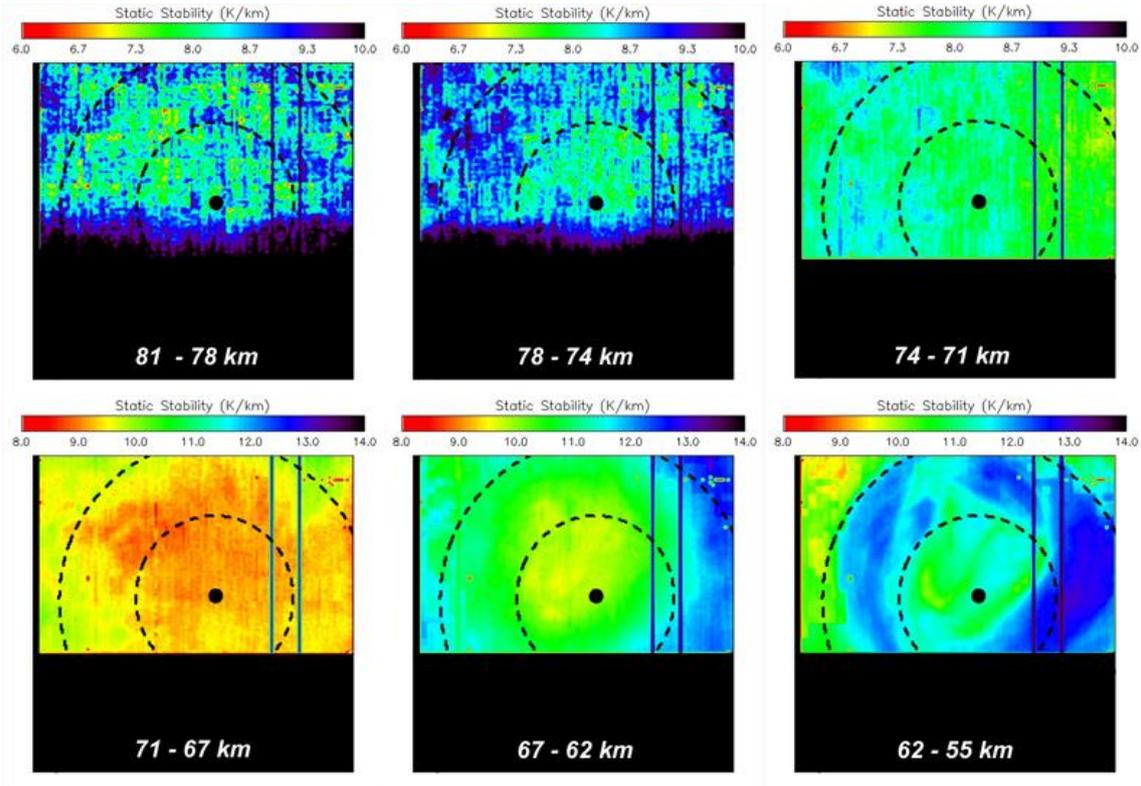

**Figure 10:** Static stability distribution on orbit 475 for several atmospheric layers (a) 81 – 78 km, (b) 78 – 74 km, (c) 74 – 71 km, (d) 71 – 67 km, (e) 67 – 62 km, and (f) 62 – 55 km. The estimated errors are 0.5 K/km, 0.4 K/km, 1.0 K/km, 0.5 K/km, 0.6 K/km and 1.7 K/km, respectively. Black dashed lines depict isolatitude lines each 10º from the south pole. The cold collar is the most stable feature between 55 and 67 km. In this altitude range the hot filaments in the vortex are the less stable features. Note the scale change between the top and bottom rows.

## 4. Discussion

The retrieved thermal profiles show the characteristic inversion layer of the cold collar from ~61 to ~69 km (see Figure 1 panel b) and overall agree with the thermal profiles shown by Haus et al. (2014) for latitudes higher than 60ºS. This is also reflected in the zonal average thermal fields obtained for each of the orbits being analyzed (Figure 7) and in the mean global thermal field (Figures 7 and 11). The cold collar is clearly distinguishable between 55 and 67 km (centered at about 100 mbar), agreeing with the results from Tellmann et al. (2009) and Haus et al. (2014). This top limit of the thermal inversion region divides the atmosphere in two, at least for the three dates considered in this analysis. The upper part is more homogeneous and has long-scale horizontal temperature differences of about 25 K over horizontal distances of ~2000 km. The lower part, on the other hand, shows more fine-scale structure, creating the vortex morphology and thermal differences of up to about 50 K over the same pressure level (Figures 4, 5, and 6). This lower part of the atmosphere is highly affected by the effects of the upper cloud deck, leading to stronger local temperature variations and larger uncertainties in the retrieval.

Zasova et al. (2007) noted that the upper clouds are inhomogeneous in the thermal infrared at high latitudes, making the level where opacity equals one to range broadly from 54 km to 70 km. Certainly, different studies reported different cloud top altitude levels. Ignatiev et al. (2009) obtained cloud top altitudes of 63 – 69 km in the polar region considering an optical depth equal to one at 1.6 µm wavelength, and Lee et al. (2012) a cloud top of 62.8 ± 4.1 km in the pole by fitting VIRTIS spectra at 4 – 5 µm. Performing retrievals of temperature profiles and cloud structure from VIRTIS-M-IR measurements, Haus et al. (2014) obtained values of 60 – 63 km for the cloud top altitude close to the south pole.

Figure 11 shows the mean cloud top altitude as a function of latitude (Equator to Pole) from observations at several wavelengths as reported by different teams. These analyses agree in that the cloud top altitude decreases towards the pole, and show the polar area as a depression region (Titov et al., 2008) with $Z_{cloud}$ values that fall into the altitude range covered in this work, i.e. 56 – 64 km. Our analysis (covering night-side data from three specific dates) also suggests that latitudes equatorwards of the cold collar show an optimal fit when values of $H_{aer}$~4 km and $Z_{cloud}$~64 km are considered. In contrast, latitudes associated with the cold collar and the vortex fit better with lower values of $H_{aer}$ and $Z_{cloud}$. The apparent $H_{aer}$ decrease agrees with the results obtained by Lee et al. (2012) and Zasova et al. (1993, 2007). However, and as noted earlier, our method is not ideally suited for the retrieval of the cloud parameters, and any conclusion about them must be taken carefully.

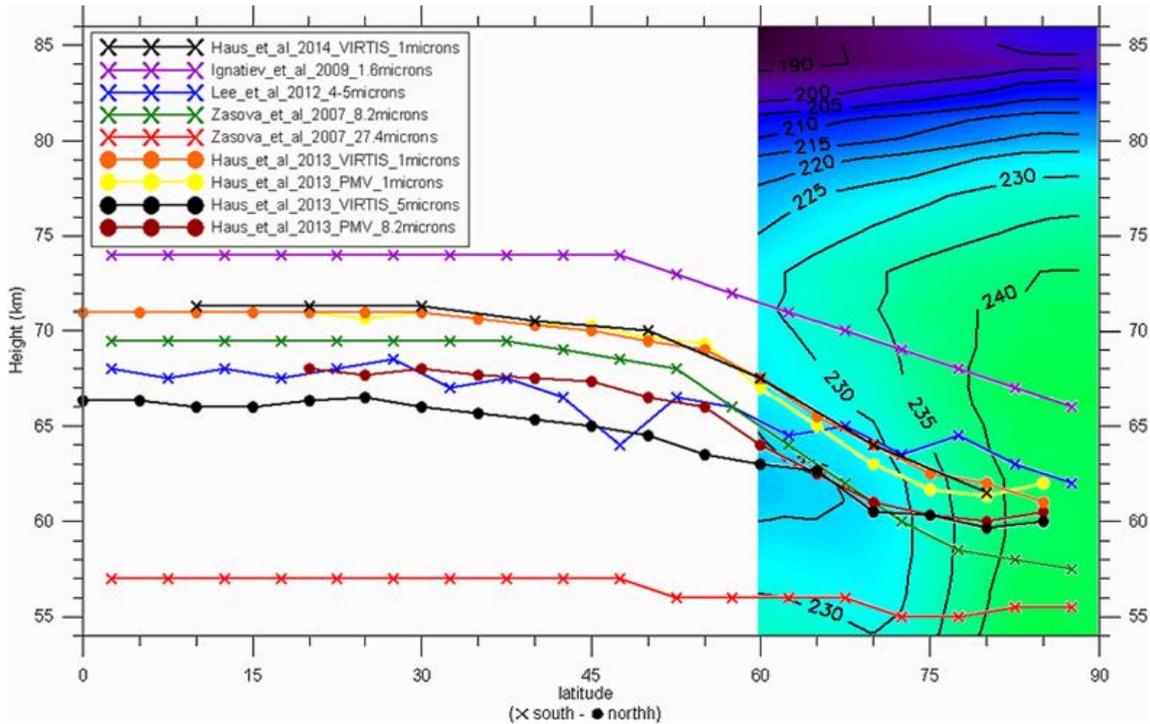

**Figure 11:** Comparison of the cloud top altitude derived from different observations over both hemispheres, superimposed to the obtained zonal averaged thermal field. Sources of the cloud top altitude: Zasova et al. (2007), Ignatiev et al. (2009), Lee et al. (2012) and Haus et al. (2013, 2014).

The South Polar Vortex morphology as seen in 5.1 µm radiance images is comparable to the structure seen in the lowest-altitude temperature maps (Figures 4, 5, and 6, panels g and h). In particular, on orbits 038 and 310, fine structure is discerned between ~55 and ~63 km and in orbit 475, the full vortex morphology appears only in the deepest level at ~55 km. This is a consequence of the dispersion in optimal $Z_{cloud}$ values throughout the polar region. The former two cases show a larger variability in the range 56 – 62 km (Figure 2), while orbit 475 is typically best fitted with a low $Z_{cloud}$. Nevertheless, the fact that in every case the 5.1 µm morphology is recovered in the thermal maps which correspond to the obtained cloud top altitude confirms that the wind fields obtained by image correlation techniques at 3.80 or 5.10 µm (Luz et al. 2011; Garate-Lopez et al., 2013) refer to the cloud top which ranges from 56 to 62 km when $Z_{cloud}$ is defined using $\lambda_* = 4\ \mu m$.

The wind field in the vortex seems uncorrelated with the cloud morphology, but the bright features observed in the 3.8 or 5.1 µm images are generally located surrounding the regions of maximum relative vorticity derived from wind measurements at the upper clouds level (see Figure 3 by Garate-Lopez et al. (2013)). Our thermal results confirm the suspicion that the hottest regions are surrounded by vorticity peaks due to the similarities between radiance and thermal structures at high pressure levels (see Figure 12).

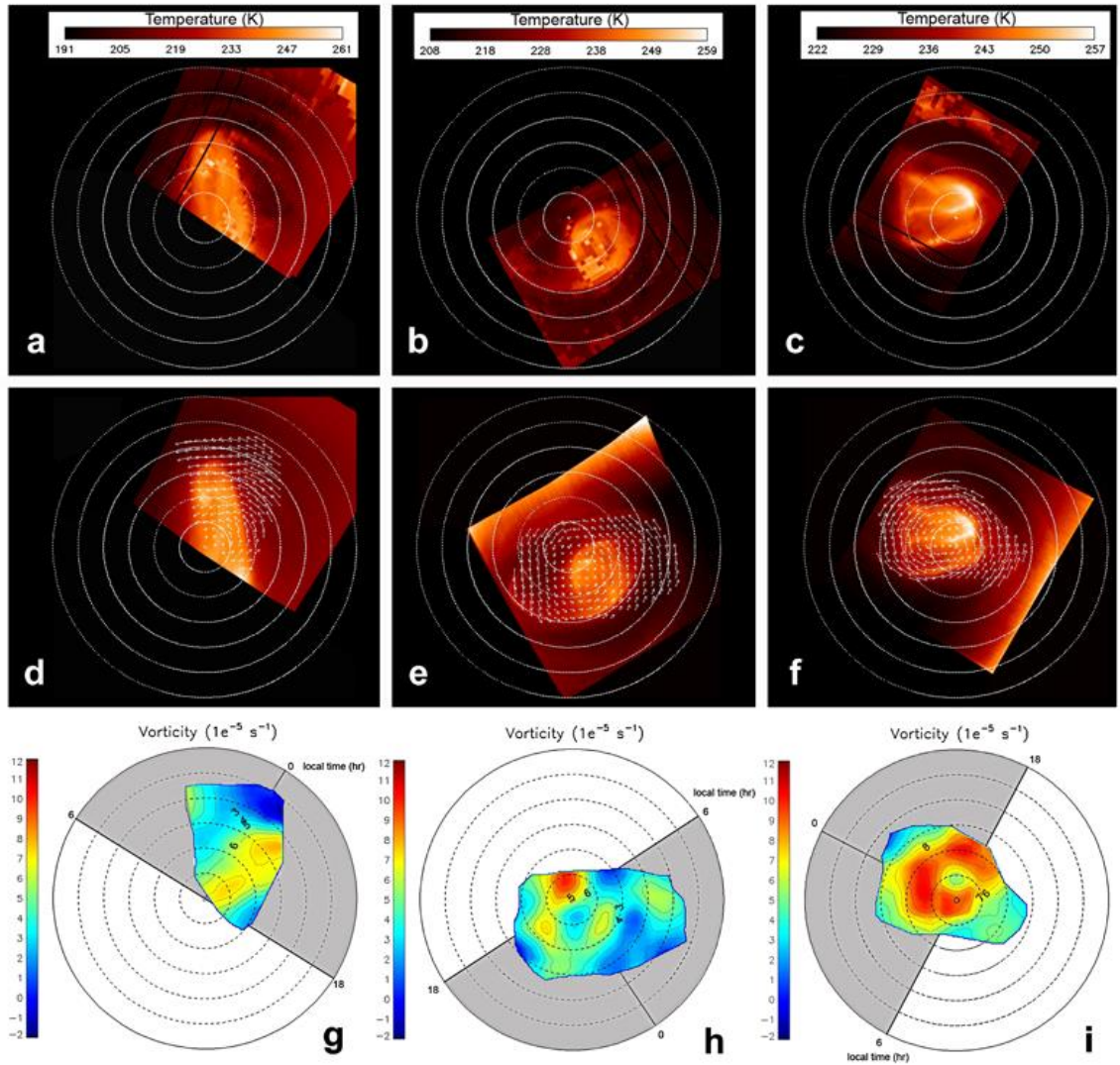

**Figure 12:** (a, b, c) Temperature maps at 360 mbar pressure level (~55 km) on orbits 038 (a), 310 (b), and 475 (c). (d, e, f) Measured wind field superimposed over the polar projections of the vortex morphology as observed at ~5 μm for each previous orbit. The largest vectors are 68 m/s, 50 m/s, and 61 m/s in panel d, e, and f, respectively. Typical wind measurement errors are 4 m/s. (g, h, i) Maps of the relative vorticity distribution derived from the wind measurements.

We suggested that the source of the vortex variability is possibly related to the location of the vortex in a baroclinic environment within a quasi-convective region embedded between the upper and lower cloud layers (Garate-Lopez et al., 2013). The presence of vertical wind shear and horizontal temperature gradients in a highly statically stable atmosphere (as reported here in the 55 - 85 km altitude range) tend to favor baroclinic instabilities. However, the current work cannot fully clarify this issue since our temperature retrieval is not sensitive to the altitude range where the static stability is low (45 – 55 km). But, since our static stability analysis above ~55 km agrees with that of Tellmann et al. (2009), who found unstable regions below this level (within the middle cloud deck), it is reasonable to presume that convective regions may also exist at polar latitudes and play an important role in the high variability of the vortex.

## 5. Summary and Conclusions

We have applied the atmospheric model described by García Muñoz et al. (2013) and a variant of the retrieval algorithm detailed in Grassi et al. (2008) to obtain thermal maps of the Venus south polar region between 55 and 85 km altitudes from VIRTIS-M-IR night-side data. These maps have been retrieved for three different dynamical configurations of the vortex whose dynamics in terms of cloud motions have been previously obtained (Garate-Lopez et al., 2013).

Both the temperature maps at different pressure levels and the zonal average thermal fields of the three orbits under investigation show that the top limit of the thermal signature from the South Polar Vortex is at ~80 km altitude. The vortex is constrained to latitudes higher than 75º by the cold collar, which is placed between altitudes ~55 km and ~67 km.

Temperature maps retrieved at 55 – 63 km show the same structures that are observed in the ~5 µm radiance images. This altitude range coincides with the optimal values of the cloud top altitude at polar latitudes and we conclude that any magnitude derived from the analysis of ~5 µm images is measured at 56 – 62 km altitude range. These values are slightly lower than previously thought (Lee et al. 2012, Haus et al. 2013). Of specific interest in the understanding of the Venus wind dynamics, the wind fields obtained by image correlation techniques using VIRTIS-M-IR night-side images at ~5 µm reflect the dynamics of the cloud top at 56 – 62 km.

Using the maps of vertically-resolved temperatures, we studied the vertical stability of the atmosphere dividing the 55 – 85 km vertical range in seven layers of approximately 4 km thickness. The layer between 62 and 67 km resulted to be the most stable in the three vortex configurations that were analyzed. The cold collar is clearly the most statically stable structure at night-time polar latitudes, while the vortex and sub-polar latitudes show lower stability values. Furthermore, the hot filaments present within the vortex exhibit even lower stability values than their surroundings. This stability distribution was to some extent expected since the cold collar is a permanent feature in the Venusian atmosphere while the vortex evolves constantly.

Although errors at altitudes below 60 km are significant because of the difficulty to probe these levels with VIRTIS data and the cloud effect, we retrieved temperatures down to 55 km, since the obtained optimal $Z_{cloud}$ values are deep. Importantly, we note that the conclusions derived from the averaged thermal fields and from the static stability structure do not change even ignoring the lowermost temperature maps.

Temperature maps retrieved in this work, together with previously measured wind fields (Garate-Lopez et al., 2013), will be used in a subsequent work to study the spatial distribution of potential vorticity in the south polar vortex for the three dynamical configurations analyzed here. In turn, that will allow us to improve our understanding of the dynamical characteristics of the vortex and its unpredictable character. The ultimate goal will be to develop a consistent model for polar vortices on Venus and the role they play in the general atmospheric circulation and Venus' superrotation.

**Acknowledgements**

We wish to thank ESA for supporting the Venus Express mission, ASI (by the contract I/050/10/0), CNES and the other national space agencies supporting the VIRTIS instrument onboard Venus Express and their principal investigators G. Piccioni and P. Drossart. This work was supported by the Spanish MICIIN projects and AYA2012-36666 with FEDER support, PRICI-S2009/ESP-1496, Grupos Gobierno Vasco IT-765-13 and by Universidad País Vasco UPV/EHU through program UFI11/55. IGL and AGM gratefully acknowledge ESA/RSSD for hospitality and access to 'The Grid' computing resources.

# References

Barstow, J. K. *et al.* Models of the global cloud structure on Venus derived from Venus Express observations. *Icarus* **217,** 542–560 (2012).

Bertaux, J.-L. *et al.* SPICAV oon Venus Express: Three spectrometers to study the global structure and composition of the Venus atmosphere. *Planetary and Space Science* **55,** 1673–1700 (2007a).

Bertaux, J.-L. *et al.* A warm layer in Venus' cryosphere and high-altitude measurements of HF, HCl, H2O and HDO. *Nature* **450,** 646–9 (2007b).

Drossart, P. *et al.* Scientific goals for the observation of Venus by VIRTIS on ESA/Venus express mission. *Planet. Space Sci.* **55,** 1653–1672 (2007).

Epele, L. N., Fanchiotti, H., García Canal, C. A., Pacheco, A. F. & Sañudo, J. Venus atmosphere profile from a maximum entropy principle. *Nonlinear Process. Geophys.* **14,** 641–647 (2007).

Garate-Lopez, I. *et al.* A chaotic long-lived vortex at the southern pole of Venus. *Nat. Geosci.* **6,** 254–257 (2013).

García Muñoz, A., Wolkenberg, P., Sánchez-Lavega, A., Hueso, R. & Garate-Lopez, I. A model of scattered thermal radiation for Venus from 3 to 5 microns. *Planet. Space Sci.* **81,** 65–73 (2013).

Grassi, D. *et al.* Retrieval of air temperature profiles in the Venusian mesosphere from VIRTIS-M data: Description and validation of algorithms. *J. Geophys. Res.* **113,** 1–12 (2008).

Grassi, D. *et al.* Thermal structure of Venusian nighttime mesosphere as observed by VIRTIS-Venus Express. *J. Geophys. Res.* **115,** 1–11 (2010).

Grassi, D. *et al.* The Venus nighttime atmosphere as observed by the VIRTIS-M instrument. Average fields from the complete infrared dataset. *J. Geophys. Res. Planets* (2014). doi:10.1002/2013JE004586

Hanel, R. A., Conrath, B. J., Jennings, D. E. & Samuelson, R. E. *Exploration of the Solar System by Infrared Remote Sensing, Second Edition*. 536 (Cambridge University Press, 2003).

Haus, R., Kappel, D. & Arnold, G. Self-consistent retrieval of temperature profiles and cloud structure in the northern hemisphere of Venus using VIRTIS/VEX and PMV/VENERA-15 radiation measurements. *Planet. Space Sci.* **89,** 77–101 (2013).

Haus, R., Kappel, D. & Arnold, G. Atmospheric thermal structure and cloud features in the southern hemisphere of Venus as retrieved from VIRTIS/VEX radiation measurements. *Icarus* **232,** 232–248 (2014).

Häusler, B. *et al.* Radio Science Investigations by VeRa Onboard the Venus Express Spacecraft. *Planet. Space Sci.* **54**, 1315–1335 (2006).

Hueso, R., Legarreta, J., García-Melendo, E., Sánchez-Lavega, A. & Pérez-Hoyos, S. The jovian anticyclone BA II. Circulation and interaction with the zonal jets. *Icarus* **203,** 499–515 (2009).

Ignatiev, N. I. *et al.* Altimetry of the Venus cloud tops from the Venus Express observations. *J. Geophys. Res.* **114,** 1–10 (2009).

Lee, Y. J. *et al.* Vertical structure of the Venus cloud top from the VeRa and VIRTIS observations onboard Venus Express. *Icarus* **217,** 599–609 (2012).

Luz, D. *et al.* Venus's Southern Polar Vortex Reveals Precessing Circulation. *Science.* **332,** 577–580 (2011).

Markiewicz, W.J. *et al.* Venus monitoring camera for Venus Express. *Planet. Space Sci.* **55**, 1701–1711 (2007).

Migliorini, a. *et al.* Investigation of air temperature on the nightside of Venus derived from VIRTIS-H on board Venus-Express. *Icarus* **217,** 640–647 (2012).

Peralta, J. *et al.* Solar migrating atmospheric tides in the winds of the polar region of Venus. *Icarus* **220,** 958–970 (2012).

Piccialli, a. *et al.* Cyclostrophic winds from the Visible and Infrared Thermal Imaging Spectrometer temperature sounding: A preliminary analysis. *J. Geophys. Res.* **113,** E00B11 (2008).

Piccioni, G. *et al.* South-polar features on Venus similar to those near the north pole. *Nature* **450,** 637–40 (2007).

Roos-Serote, M. C. *et al.* Thermal Structure & Dynamics Atmosphere Venus 70-90km From Galileo NIMS. *Icarus* **114,** 300–309 (1995).

Sánchez-Lavega, A., An Introduction to Planetary Atmospheres, *Taylor & Francis*, 2011. **587pp**.

Schofield, J. T. & Diner, D. J. Rotation of Venus's Polar Dipole. *Nature* **305,** 116–119 (1983).

Seiff, A. Thermal structure of the atmosphere of Venus. In *Venus*, eds. D. M. Hunten, L. Colin, T. M. Donahue and V. I. Moroz (Tucson: Univ. of Arizona Press), pp. 215-279.

Seiff, A. *et al.* Measurements of thermal structure and thermal contrasts in the atmosphere Venus and related dynamical observations: Results from the four Pioneer Venus probes. *J. Geophys. Res.* **85**, 7903-7933 (1980).

Stamnes, K., Tsay, S. C., Jayaweera, K. & Wiscombe, W. Numerically stable algorithm for discrete-ordinate-method radiative transfer in multiple scattering and emitting layered media. *Applied Optics* **27**, 2502-2509 (1988).

Suomi, V. E. & Limaye, S. S. Venus: Further Evidence of Vortex Circulation. *Science.* **201,** 1009–1011 (1978).

Svedhem, H. *et al.* Venus Express—The first European mission to Venus. *Planet. Space Sci.* **55**, 1636–1652 (2007).

Taylor, F. W. *et al.* Structure and Meteorology of the Middle Atmosphere of Venus: Infrared Remote Sensing From the Pioneer Orbiter. *J. Geophys. Res.* **85,** 7963–8006 (1980).

Tellmann, S., Pätzold, M., Häusler, B., Bird, M. K. & Tyler, G. L. Structure of the Venus neutral atmosphere as observed by the Radio Science experiment VeRa on Venus Express. *J. Geophys. Res.* **114,** 1–19 (2009).

Titov, D. V *et al.* Atmospheric structure and dynamics as the cause of ultraviolet markings in the clouds of Venus. *Nature* **456,** 620–3 (2008).

Titov, D. V. *et al.* Morphology of the cloud tops as observed by the Venus Express Monitoring Camera. *Icarus* **217,** 682–701 (2012).

Wilson, C. F., Guerlet, S., Irwin, P. G. J., Tsang, C. C. C., Taylor, F. W., Carlson, R. W., … Piccioni, G. (2008). Evidence for anomalous cloud particles at the poles of Venus. *J. of Geophys. Res.,* **113**, 1–12. doi:10.1029/2008JE003108

Zasova, L. V, Moroz, V. I., Esposito, L. W. & Na, C. Y.. $SO_2$ in the Middle Atmosphere of Venus: IR Measurements from Venera-15 and Comparison to UV Data. *Icarus* **105,** 92–109 (1993).

Zasova, L. V, Khatountsev, I. A., Moroz, V. I. & Ignatiev, N. I.. Structure of the Venus middle atmosphere: Venera 15 Fourier Spectrometry data revisited. *Adv. Sp. Res.* **23,** 1559–1568 (1999).

Zasova, L. V, Ignatiev, N. I.. Khatountsev, & Linkin, V.. Structure of the Venus atmosphere. *Planet. Space Sci.* **55,** 1712–1728 (2007).